\documentstyle[11pt,newpasp,twoside,epsf]{article}
\begin{document}

\title{The Wisconsin H-Alpha Mapper Northern Sky Survey}
\author{L. M. Haffner}
\affil{Department of Astronomy, University of Wisconsin--Madison, 475
  North Charter St., Madison, WI 53703}

\begin{abstract}
  The Wisconsin H-Alpha Mapper (WHAM) has completed a survey in Balmer
  $\alpha$ of the entire Northern sky above declination $-30\deg$.
  This survey provides the first calibrated, velocity-resolved map of
  the H$\alpha$ emission from the Galaxy. With one-degree spatial
  resolution, 12 km s$^{-1}$ velocity resolution, and sensitivity to
  features as faint as 0.1 R (EM $\sim$ 0.2 cm$^{-6}$ pc), this survey
  provides the deepest maps of the ionized content of the Galaxy to
  date. In addition to probing the detailed kinematic structure of the
  Warm Ionized Medium and the vertical structure of the ionized
  content in spiral arms, initial results include the discovery of
  several faint, extended ($d > 1\deg$) H {\sc ii} regions and the
  first map of the ionized component of an intermediate velocity
  cloud.
\end{abstract}

\section{Introduction}
\label{sec:intro}

Past studies have shown the Warm Ionized Medium (WIM) to be a
significant component of the ISM, especially in the halos of disk
galaxies (Reynolds 1993; Hoopes, Walterbos, \& Rand 1999; Rand 1996;
Rossa \& Dettmar 2000; Dettmar 1998). In the Milky Way, the WIM has a
mass surface density about one-third that of H I, an ionization power
requirement equal to the supernovae rate, and a characteristic scale
height of 1 kpc (Reynolds 1993; Haffner, Reynolds, \& Tufte 1999).
Although the details of how the WIM is ionized and heated are not well
understood yet, early-type stars seem to be the biggest contenders for
providing the bulk of the ionizing photons. Dove, Shull, \& Ferrara
(2000); Dove \& Shull (1994); and Miller \& Cox (1993) have studied
how O-star radiation can leak from the plane to ionize the large
expanse occupied by the WIM. However, we will show below that B-stars
may also be a significant contributer to the halo ionization field.
The problem of Lyman continuum propagation through neutral gas is
somewhat easier if these more widely distributed stars can contribute.

Aided by modern detector technology, several groups have initiated
intensive programs to characterize the global details of the ionized
content of galaxies (also called Diffuse Ionized Gas---DIG). In
addition to the velocity-resolved all-sky survey described here, three
large-area imaging surveys are also in progress. A southern survey
with arc-minute resolution has been completed by Gaustad, McCullough,
\& Van Buren (1996; McCullough, these proceedings). Dennison,
Simonetti, \& Topasna (1998) have been imaging the northern sky with
similar resolution. A high-resolution ($\sim 1\arcsec$) survey of the
Galactic plane and Magellanic clouds is also underway by Parker \&
Phillips (1998).

Here we describe the results of the Wisconsin H-Alpha Mapper (WHAM)
Northern Sky Survey, the first deep ($I_{\rm H\alpha} \sim 0.1$ R; 1 R =
$10^6/4\pi$ photons cm$^{-2}$ s$^{-1}$ ster$^{-1}$), velocity-resolved
survey of the WIM in our Galaxy.

\section{The WHAM Survey}
\label{sec:survey}

The northern portion of the WHAM sky survey contains nearly 37,300
spectra with 12 km s$^{-1}$ resolution, each representing the
spatially integrated emission from a one-degree patch on the sky. The
survey covers the northern sky above $\delta \ge -30\deg$ with a beam
spacing of $\Delta b = 0\fdg85$ and $\Delta \ell = 0\fdg98 / \cos b$.
Each spectrum covers a 200 km s$^{-1}$ spectral interval centered near
the Local Standard of Rest (LSR). With our spectral resolution, we are
able to remove the bright geocoronal H$\alpha$ line and several very
faint atmospheric lines present in every spectrum. These cleaned
spectra allow us to detect very faint extended sources (I$_{H\alpha}$
= 0.1 R; EM $\sim 0.2$ cm$^{-6}$ pc) and will allow us to absolutely
calibrate the WHAM survey. More details of the WHAM instrument and the
survey strategy can be found in Tufte (1997) and Haffner (1999).

\begin{figure}[tbp]
  \begin{center}
   \plotfiddle{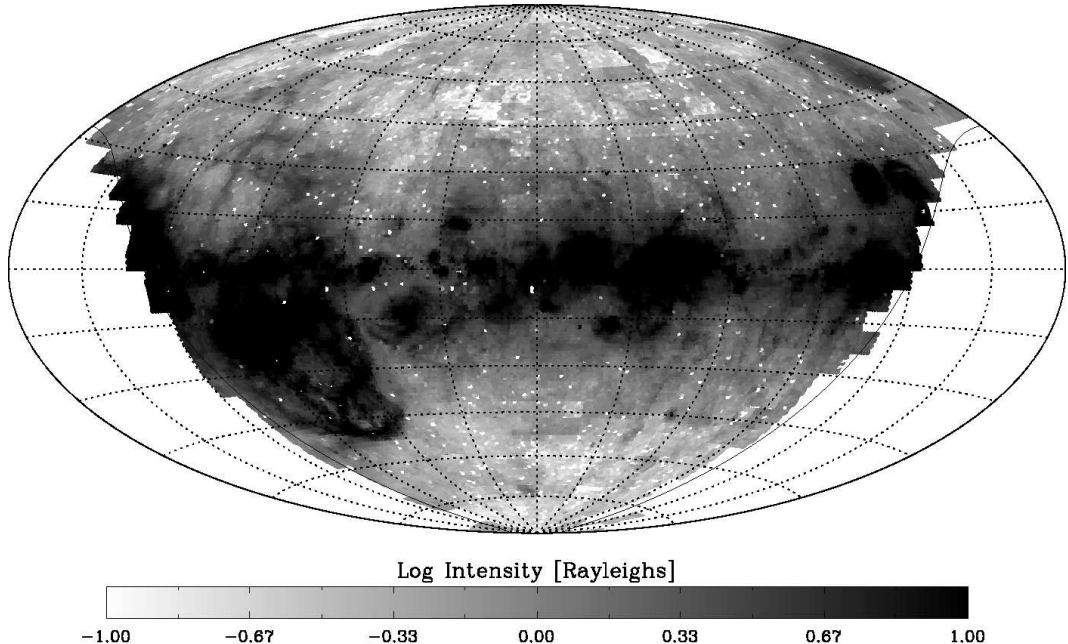}{2.8in}{90}{55}{55}{220}{-36}
    \caption{The WHAM H$\alpha$ Sky Survey. Over 37,100 beams
      are plotted on a Hammer-Aitoff Galactic projection centered at
      $\ell = 120\deg$. Grayscale color represents the total
      integrated H$\alpha$ intensity within a one-degree pointing. For
      display purposes, the intensity scale is limited from 0.1 to 10
      R (EM $\sim$ 0.2 to 20 cm$^{-6}$ pc). Dotted lines are grid
      lines spaced 30\deg\ apart in longitude and 15\deg\ apart in
      latitude. The solid line is $\delta = -30\deg$, the limit of the
      northern survey. White points are beams contaminated by very
      bright stars ($V \la 6$ mag) with significant H$\alpha$
      absorption.}
    \label{fig:allsky}
  \end{center}
\end{figure}

The nearly completed (99\%) northern sky survey is displayed in
Figure~\ref{fig:allsky}. As can been seen, very few regions exist
where we do not detect at least 0.1 R of H$\alpha$ emission. Much of
the intensity variation at high latitudes is real although the final
phase of intensity calibrations has not yet been applied to the data
presented here. To illustrate the power of this new survey we present
two new discoveries here: the first map of the ionized component of an
Intermediate Velocity Cloud (IVC) and the detection of several faint,
diffuse H {\sc ii} regions.

\section{The Ionized Component of Complex K}
\label{sec:compk}

\begin{figure}[tbp]
  \begin{center}
    \plotone{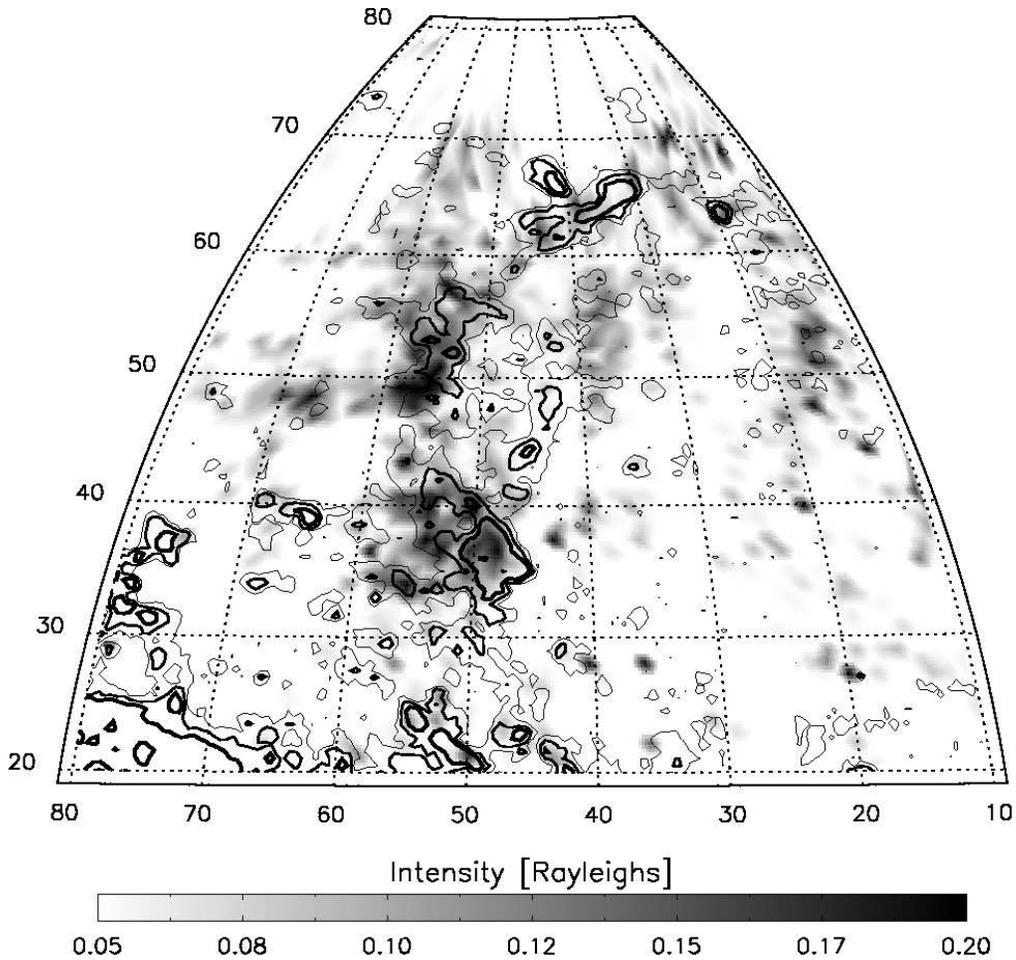}
    \caption{The Ionized and Neutral Components of Complex K.
      Integrated H$\alpha$ emission from $v_{LSR} = -95$ to $-60$ km
      s$^{-1}$ is presented in the grayscale image. H {\sc i} column
      densities (integrated over the same velocity range) of 5, 10,
      and $20 \times 10^{18}$ cm$^{-2}$ from Hartmann \& Burton (1997)
      are displayed as contours.}
    \label{fig:compk}
  \end{center}
\end{figure}

Wakker (2000) recently separated out the H {\sc i} intermediate
velocity Complex K (previously classified with Complex C) centered
near $\ell = 50\deg$, $b = +50\deg$. Figure~\ref{fig:compk} shows
an image of the H$\alpha$ emission from WHAM and contours of 21 cm
emission from the Leiden-Dwingeloo H {\sc i} survey (Hartmann \& Burton
1997) toward Complex K.

The H$\alpha$ boundary matches the $N_H = 5 \times 10^{18}$ cm$^{-2}$
contour quite closely over most of the cloud; however, the peaks in
each emission component do not correspond as well. The H$\alpha$
intensity over the face of the cloud only varies by about a factor of
2: 0.1 to 0.2 R or EM $\approx$ 0.2 to 0.4 cm$^{-2}$ pc. The total
$N_{H\;I}\ (-100 < v_{LSR} < +100)$ over the face of the cloud varies
from about 1 to $4 \times 10^{20}$ cm$^{-2}$, resulting in an
extinction correction to the observed H$\alpha$ flux of at most $25\%$
if all this neutral material is actually between us and the cloud.

As reviewed by Wakker (2000), little absorption line data directly
toward Complex K exists at this time and thus little information about
its distance. There is a firm upper limit established though since the
cloud has been seen in absorption toward nearby M13 ($\ell = 59\deg, b
= +41\deg$). Carretta, et al.\ (2000) have recently calculate the
distance modulus to M13 to be 14.44 mag, setting an upper limit on the
distance of the portion of complex K near M13 at 7.7 kpc.

If the cloud is ionized by an external photoionizing source the
incident Lyman continuum flux on the cloud surface can be estimated
(see Tufte, Reynolds, \& Haffner 1998). Ignoring the complexity of
geometry but applying the maximum extinction correction above, we find
that the incident flux of ionizing photons, $\phi < 5.25 \times 10^5$
photons cm$^{-2}$ s$^{-1}$ for the brightest H$\alpha$ regions in
Complex K.  Bland-Hawthorn \& Maloney (1999) have produced a model of
the ionizing flux in the Galactic halo. We find that our observed
H$\alpha$ intensity is consistent with that expected from their model
in the direction of Complex K for distances $< 8$ kpc from the sun
(Bland-Hawthorn 2000, private communication).

\section{Faint, Diffuse H {\sc ii} Regions}
\label{sec:hii}

\begin{figure}[tbp]
  \begin{center}
    \plotone{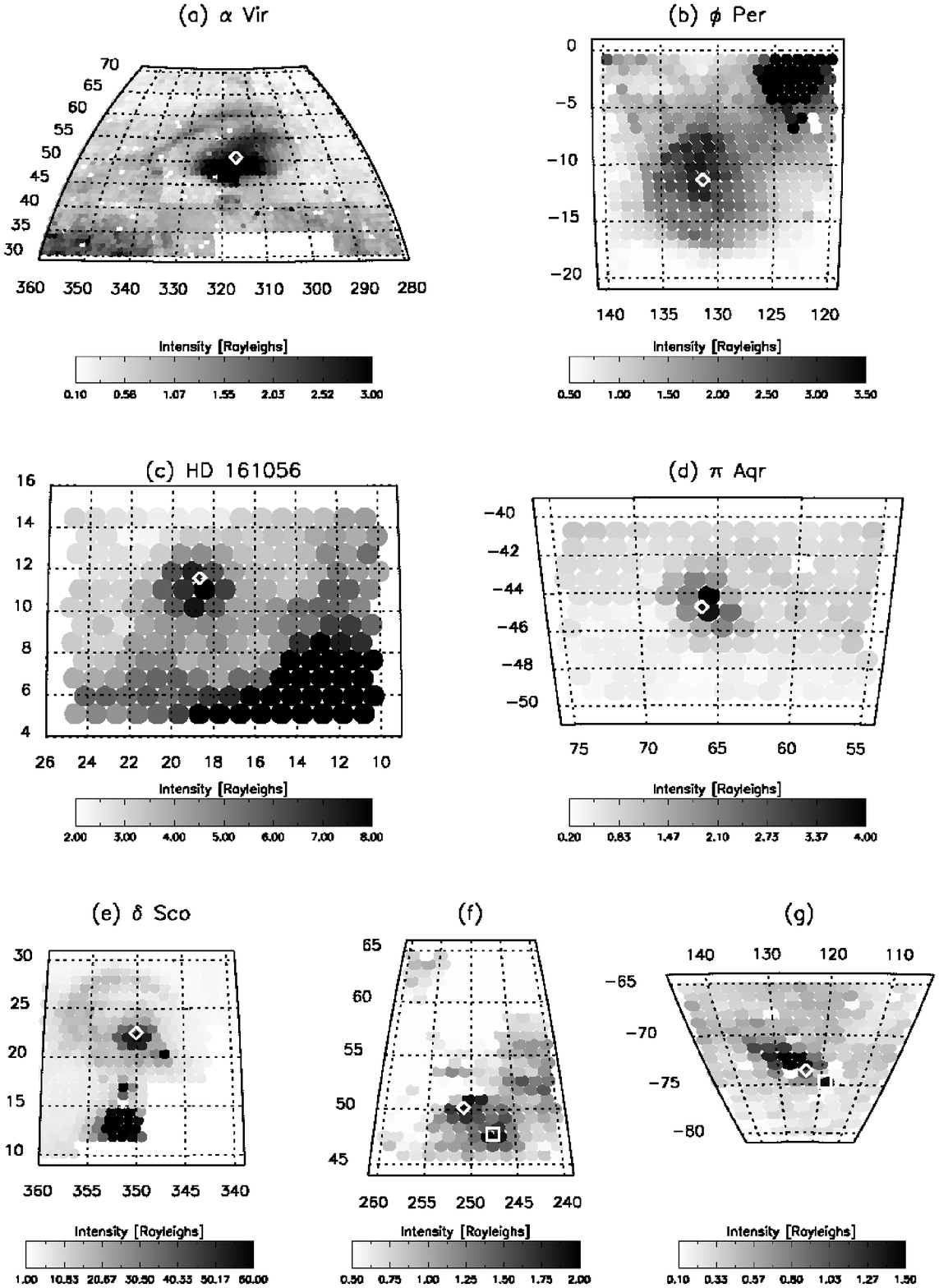}
    \caption{Faint, Diffuse H {\sc ii} Regions.
      Each of the above images has been extracted from the WHAM survey
      and, except for (b), represents the total integrated H$\alpha$
      intensity between $-100$ and $+100$ km s$^{-1}$. In (b), the
      integration range is limited to a region close to the H {\sc ii}
      emission line ($-10$ to $+10$ km s$^{-1}$) since emission from
      the more distant Perseus arm gas (centered near $-40$ km
      s$^{-1}$) would obscure the H {\sc ii} region. The diamond in
      each figure denotes the location of the likely source listed in
      the figure title. In (f) the two candidate sources are
      $(\diamond)$~PG1047+003 and $(\Box)$~WD1034+001. In (g) they are
      $(\diamond)$~PHL 6783 and $(\Box)$~NGC 246 (see text).}
    \label{fig:hii}
  \end{center}
\end{figure}

As noted above, the problem of ionizing the WIM layer becomes somewhat
easier if O stars in the plane are aided by another, more widely
distributed population. Reynolds (1985) mapped a large, low-density
H~{\sc ii} region around $\alpha$ Vir (Spica), a B1III-IV + B2 V
system. In Figure~\ref{fig:hii}, we present a full map of the Spica
H~{\sc ii} region as well as several other candidate H~{\sc ii}
regions discovered in the WHAM survey.

\begin{table}[tbp]
  \begin{center}
    \caption{H II Region Characteristics}
    \scriptsize
    \begin{tabular}{lcccccc}
      \tableline
      Source\tablenotemark{a} & $\ell$ & $b$ & Distance\tablenotemark{b} &
      Spectral Type & Diameter & $L_{LC}$ \\
      
      & $[\deg]$ & $[\deg]$ & [pc] &
      & [\deg/pc] & [photons s$^{-1}$] \\
      \tableline

      \\
      (a) $\alpha$ Vir & 316.11 & +50.84 & 80 &
      B1III-IV + B2V & 16\tablenotemark{c} / 22 & $1.23 \times 10^{46}$ \\

      (b) $\phi$ Per & 131.32 & $-11.33$ & 220 &
      B0.5Ve + sdO & 13 / 50 & $1.09 \times 10^{47}$ \\

      (c) HD 161056 & 18.67 & +11.58 & 427 &
      B1.5V & 4 / 30 & $2.77 \times 10^{46}$ \\

      (d) $\pi$ Aqr & 66.01 & $-44.74$ & 338 &
      B1Ve & 2.5 / 30 & $1.29 \times 10^{46}$ \\

      (e) $\delta$ Sco & 350.1 & +22.49 & 123 &
      B0.2IV & 3\tablenotemark{d} / 13  & $5.39 \times 10^{46}$ \\

      (f) PG1047+003 & 250.85 & +50.17 & -- &
      sdB & 1.5 & -- \\

      (f) WD1034+001 & 247.55 & +47.75 & -- &
      DO & 1.5 & -- \\

      (g) PHL 6783 & 123.64 & -73.53 & -- &
      sdO & 2 & -- \\

      (g) NGC 246 & 118.86 & -74.71 & 600: &
      Op & 2 & -- \\
      \\

      \tableline \tableline
    \end{tabular}
    
    \tablenotetext{a}{The letter in parentheses refers to a particular
      H~{\sc ii} region in Figure~\ref{fig:hii}.}
    
    \tablenotetext{b}{{\it Hipparcos} (Perryman et al.\ 1997).}
    
    \tablenotetext{c}{The emission region around $\alpha$ Vir is
      noticeably non-circular (even at our one-degree scale), most
      likely due to a ridge of H {\sc i} that forms the southern
      boundary of the nebula (see Fejes 1974). This number is a rough
      estimate of its extent. Although its shape and location suggest
      association, the faint bar of emission seen to the northeast of
      the main nebula is not included in this calculation. If
      included, the extra emission would increase the measured
      $L_{LC}$ by about 10\%.}
    
    \tablenotetext{d}{If the bar of emission seen to the north and
      east of the main nebula is included, $L_{LC}$ increases by about
      25\%.}
    
    \label{tab:hii}
  \end{center}
\end{table}

The likely sources for these regions are summarized in
Table~\ref{tab:hii}. We estimate the physical diameter of each region
when the distance to the source is known. In these cases, the final
column lists the minimum Lyman continuum luminosity needed to sustain
the observed H$\alpha$ flux:
\begin{equation}
  \label{eq:lc}
  L_{LC} = \frac{4 \pi d^2}{\epsilon}
  \int I_{H\alpha}\,e^{\tau_{H\alpha}}\,d\Omega,
\end{equation}
where $d$ is the distance to the region, $I_{H\alpha}$ is the
$H\alpha$ intensity emitted by the solid angle $d\Omega$, $\epsilon$
is the fraction of $H\alpha$ photons produced per Lyman continuum
photon (0.47 at 8000 K; Martin 1988), and $e^{\tau_{H\alpha}}$ is the
extinction correction factor. Background emission from the WIM
(estimated from outside the region) is subtracted so that the
$I_{H\alpha}$ used in this calculation is (presumably) emission only
from the H {\sc ii} region.

We emphasize that our estimate of $L_{LC}$ is a lower limit for two
reasons.  First, although many of these regions appear to be circular,
we cannot be sure that all the Lyman continuum radiation from the
source is absorbed by the local gas forming the H~{\sc ii} region.
Second, we do not apply an extinction correction here ({\it i.e.}\ 
$e^{\tau_{H\alpha}} = 1$). An ongoing WHAM survey of H$\beta$ will
directly provide the extinction correction in the future.

The H {\sc ii} region displayed in Figure~\ref{fig:hii}g deserves
special mention. Current catalogs list no candidate source near the
center of this region. The only two reasonable candidates within a
5\deg\ radius of the center of the region are PHL 6783 (sdO) and NGC
246 (Op; PN). On the far side of the ionized nebula from these sources
($\ell = 115$ to 140, $b = -72$ to $-68$) sits an H\ {\sc i} cloud
whose velocity matches the H$\alpha$ emission. We interpret this
seeming coincidence as evidence that one of these sources (most likely
PHL 6783 from the geometry) is ionizing the face of this H\ {\sc i}
cloud.

However, the picture may be even a bit more interesting. The proper
motion of PHL 6783 ($\mu_\alpha = -31$ mas yr$^{-1}$, $\mu_\delta =
-31$ mas yr$^{-1}$; $\mu_\ell = -99$ mas yr$^{-1}$, $\mu_b = -32$ mas
yr$^{-1}$; H{\o}g et al. 2000) indicates that it passed through the
peak H$\alpha$ of the region approximately $1.5 \times 10^5$ years
ago. The central star of NGC 246 also has a proper motion ($\mu_\alpha
= -18$ mas yr$^{-1}$, $\mu_\delta = -10$ mas yr$^{-1}$; $\mu_\ell =
-63$ mas yr$^{-1}$, $\mu_b = -10$ mas yr$^{-1}$; H{\o}g et al. 2000)
that suggests it was just a degree south of this peak $6 \times 10^5$
years ago. The H {\sc ii} region is slightly elongated along these
proper motion vectors (both of which also eventually intersect the H\ 
{\sc i} cloud) and is almost directly aligned with that of PHL 6783.
With such proper motion vectors, there is a possibility that the H\ 
{\sc ii} region is no longer being actively ionized. In this slightly
more complicated scenario, PHL 6783 still seems to be the best
candidate for the source since its proper motion vector direction and
length are best suited to have recently ionized the region.

\section{Summary}
\label{sec:summary}

WHAM has completed a full spectral survey of the northern sky ($\delta
> -30\deg$) in H$\alpha$. Similar to the neutral component,
filamentary substructure abounds in the ionized halo of the Galaxy.
Exciting results from the faint H$\alpha$ sky are abundant and we have
presented two specific examples here. We have presented the first
ionized map of an HVC or IVC. Complex K contains an ionized component
with $I_{H\alpha} \approx 0.1$ -- $0.2$ R that follows the shape of
the neutral gas extremely well. We have also presented several new,
faint H\ {\sc ii} regions, many of which appear to be ionized by early
B stars. In addition, one of these regions is very likely an example
of an isolated H\ {\sc i} cloud being ionized by a nearby sdO star.

\acknowledgements

This research has made substantial use of the SIMBAD database,
operated at CDS, Strasbourg, France. WHAM is supported by the National
Science Foundation through grant AST 96-19424.

\end{document}